\documentclass[aip,graphicx]{revtex4-1}

\usepackage{graphicx}
\usepackage{bm}

\draft 

\begin{document}


\title{Coherent Diffractive Imaging Using Randomly Coded Masks} 



\author{Matthew H. Seaberg}
\email[]{seaberg@slac.stanford.edu}
\affiliation{CNRS \& D.I., UMR 8548, \'{E}cole Normale Sup\'{e}rieure, 45 Rue d'Ulm, 75005 Paris, France}
\affiliation{SLAC National Accelerator Laboratory, 2575 Sand Hill Road, Menlo Park, CA 94025, USA}

\author{Alexandre d'Aspremont}
\affiliation{CNRS \& D.I., UMR 8548, \'{E}cole Normale Sup\'{e}rieure, 45 Rue d'Ulm, 75005 Paris, France}

\author{Joshua J. Turner}
\affiliation{SLAC National Accelerator Laboratory, 2575 Sand Hill Road, Menlo Park, CA 94025, USA}


\date{\today}

\begin{abstract}
Coherent diffractive imaging (CDI) provides new opportunities for high resolution X-ray imaging with simultaneous amplitude and phase contrast. Extensions to CDI broaden the scope of the technique for use in a wide variety of experimental geometries and physical systems. Here, we experimentally demonstrate a new extension to CDI that encodes additional information through the use of a series of randomly coded masks. The information gained from the few additional diffraction measurements removes the need for typical object-domain constraints; the algorithm uses prior information about the masks instead. The experiment is performed using a laser diode at 532.2~nm, enabling rapid prototyping for future X-ray synchrotron and even free electron laser experiments. Diffraction patterns are collected with up to 15 different masks placed between a CCD detector and a single sample. Phase retrieval is performed using a convex relaxation routine known as ``PhaseCut" followed by a variation on Fienup's input-output algorithm. The reconstruction quality is judged via calculation of phase retrieval transfer functions as well as by an object-space comparison between reconstructions and a lens-based image of the sample. The results of this analysis indicate that with enough masks (in this case 3 or 4) the diffraction phases converge reliably, implying stability and uniqueness of the retrieved solution.
\end{abstract}


\maketitle 

\section{Introduction}
The first demonstration of X-ray coherent diffractive imaging (CDI), predicted in 1952 \cite{Sayre1952} but not realized until 1999 \cite{Miao1999}, has sparked a revolution in X-ray imaging techniques. Since the initial demonstration, numerous variations and extensions to the basic technique have resulted in a wide variety of materials science and biological applications \cite{Chapman2010,Miao2015}. CDI replaces imaging optics with an iterative phase retrieval algorithm that solves for the missing phase of the measured diffraction intensity, resulting in an aberration-free imaging system capable of producing diffraction-limited images. In addition, because CDI retrieves the complex exit surface wave due to the interaction of light with an object, this technique offers simultaneous, quantitative amplitude and phase contrast. Because of these attractive qualities, CDI is a promising technique for ultrafast imaging studies of functional materials at X-ray free electron lasers (FELs).

Despite the advantages of CDI, the basic technique suffers from limitations due to the constraints required to ensure that the algorithm will converge. Constraints are commonly enforced on the phase and support of the object to improve convergence, restricting the thickness and transverse extent of the object, respectively. Thus new techniques are needed to enable widespread use of CDI at X-ray FELs. Ptychography, a recent enhancement to CDI, removes these constraints and enables imaging of extended objects with no phase limitations \cite{Thibault2008,Maiden2009}. It involves collecting many diffraction patterns as the object is scanned across the beam; enforcing object consistency in the overlapping regions of adjacent scan positions replacing the usual object-domain constraints. The result is a very powerful technique enabling the high quality image reconstruction of a wide variety of objects in transmission \cite{Wilke2012,Hruszkewycz2013}, reflection \cite{Harada2013,Seaberg2014}, and 3D modalities \cite{Maiden2012a,Guizar-Sicairos2015}. However, its application to FELs especially for ultra-fast imaging, remains a significant challenge and different methods are needed to take full advantage of the coherent properties of these types of sources.

Here we present an experimental demonstration of a new technique that, similar to ptychography, removes the need for phase and support constraints . This method is based on recent convex relaxation algorithms that encode observations differently \cite{Candes2013,Fogel2013,Chen2015,Waldspurger2015}. It is an extension to CDI that utilizes a series of known, randomly coded masks to encode additional information into the measured diffraction patterns. The experiments are performed in the visible region of the spectrum. Based on the findings here, this new technique is expected to be well suited for use in the X-ray region at free electron laser (FEL) sources due to the small number of measurements required for retrieving high quality images.

\section{Experiment}

\begin{figure}[htb]
	\includegraphics[width=11cm]{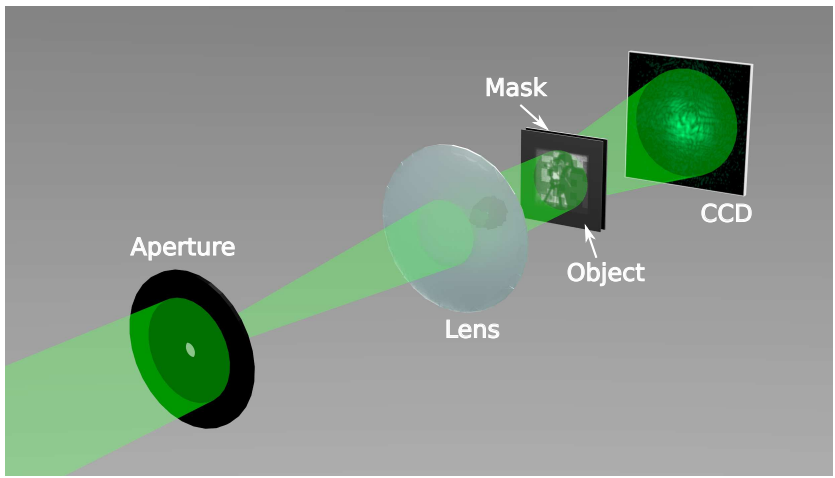}
	\caption{Schematic of the experiment. The laser is first spatially filtered using a 25~$\mu$m diameter aperture and collimated with a 10~cm focal length lens. The resulting beam illuminates the object, which is followed immediately by the randomly coded mask. The intensity pattern of the scattered light is measured on a CCD detector placed 132~mm beyond the mask.}
	\label{schematic}
\end{figure}

The experiment was performed using a laser diode at 532.2~nm wavelength. This visible-light configuration enables rapid prototyping of the technique and allows for refinements, both numerically and regarding instrumentation. A schematic of the experiment is shown in Fig.~\ref{schematic}.

\begin{figure}[htb]
	\includegraphics[width=13.5cm]{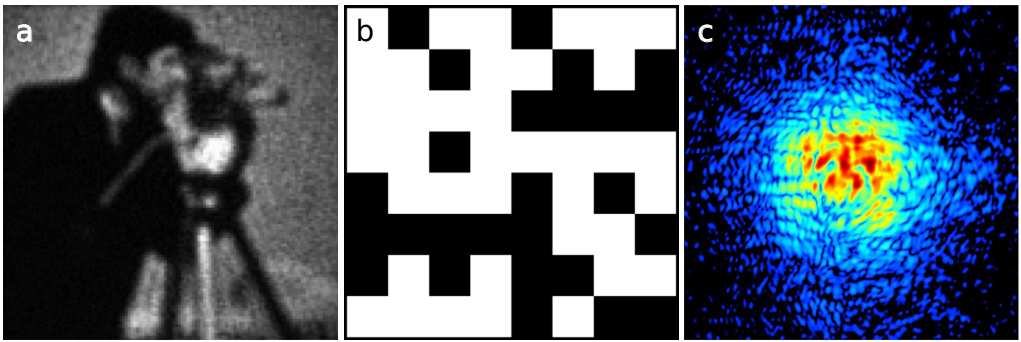}
	\caption{(a) Lens-based image of the pattern used as the unknown object in this experiment, with $\approx 10 \ \mu{\rm m}$ resolution. (b) An example of a randomly coded mask design. The full width of the mask is 1.13~mm, so that each square feature has width $141 \ \mu{\rm m}$. (c) The measured diffraction pattern due to the pattern in (a) combined with the mask in (b).}
	\label{DataFig}
\end{figure}

The laser is first sent through a 25~$\mu$m diameter aperture to spatially filter the laser beam. Following the aperture, a 10~cm focal length lens is placed 10~cm downstream of the aperture in order to collimate the transmitted light. The resulting beam has an approximate diameter of 2~mm. The sample is placed downstream of the collimating lens so that it is illuminated with a planar wavefront. The sample consists of a common test image patterned on a grayscale, 35~mm projector film (Gamma Tech). A lens-based image of the sample is shown in Fig.~\ref{DataFig}(a). A variety of random masks, also patterned on 35~mm projector film (see Fig.~\ref{DataFig}(b) for an example), are placed 1~mm downstream of the sample. Each square mask occupies an area of 1~mm$^2$, with 4~mm of opaque film separating adjacent masks, so that only one mask is illuminated by the beam at a time. Finally, a CCD detector (Allied Vision Manta G-201) collects the scattered light 132~mm downstream of the masks. The oversampling ratio, $\Omega$, in this case is $\Omega>12$, so that the diffraction pattern meets the Shannon sampling criterion \cite{Shannon1949}. Due to this geometry, the diffraction pattern is measured in the near field, and is characterized by a Fresnel number of 18. The Fresnel number is defined as $D^2/\lambda z$, where $D$ is the diameter of the object, $\lambda$ is the wavelength, and $z$ is the distance between the mask and the detector. The numerical aperture of the system is 0.017 and supports a resolution of $15\ \mu{\rm m}$, which is well below the minimum feature size of the masks ($141\ \mu{\rm m}$).

\begin{figure}
	\includegraphics[width=9cm]{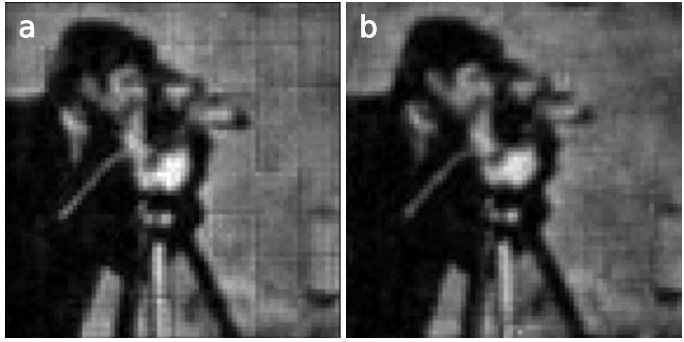}
	\caption{(a) Retrieved image using diffraction patterns from 15 different random masks, with each mask centered at the same position during the measurement. Note the square-shaped artifacts in the image. (b) Retrieved image using the same 15 masks, but with each mask position offset by a random, known amount.}
	\label{Results}
\end{figure}

The mask film is mounted on a 2-dimensional (2D) translation stage so that the masks can be switched between exposures. Mask alignment is performed by imaging the mask onto the detector through the insertion of a lens between the mask and the detector, and recording the 2D stage position for each mask when aligned to a specific region on the detector. Future improvements to the algorithm are predicted to handle errors in the relative positioning of each mask. Diffraction patterns from up to 15 different masks are collected so that the dependence of image quality can be studied as a function of the number of diffraction patterns used in the phase recovery algorithm. An example of a measured diffraction pattern is shown in Fig.~\ref{DataFig}(c). Prior to phase retrieval, the patterns are binned by a factor of 4 (so that $\Omega \approx 3$) in order to increase the speed of the algorithm. A reconstruction using all 15 masks is shown in Fig.~\ref{Results}(a). This reconstruction was performed using the ``PhaseCut'' algorithm \cite{Fogel2013}, refined by Fienup's input-output algorithm \cite{Fienup1982}. Although the PhaseCut algorithm was designed to be used in the far field, the near-field diffraction measurement does not seem to present a problem as long as the quadratic phase from the Fresnel diffraction integral is used to initialize the phase in the sample plane \cite{Quiney2006}. Explicitly, the phase at the sample plane, $\phi(\rho)$, is initialized to be $\phi(\rho) = \pi \rho^2/\lambda z$, where $\rho$ refers to the radial coordinate in the sample plane and $z$ is the distance between the sample and detector. Uniformly distributed random numbers in the range $[-\pi/2, \pi/2]$ are added to this initial guess, and the magnitude in the sample is initialized to unity. The input-output algorithm is found to reliably converge on its own as well, likely due to the increased strength of the Fourier modulus constraint in the near field \cite{Abbey2008}.

For this dataset, all 15 masks were centered in the same physical location, resulting in the square-shaped artifacts seen in the reconstruction. To remedy this, known random shifts to the mask positions were introduced for a subsequent dataset, with the result shown in Fig.~\ref{Results}(b). Since in this case the squares making up the masks were no longer aligned to each other, this technique removes the unwanted artifacts to a large extent. It should be mentioned that the mask structure and how it interacts with the illumination wavelength must be accurately known to length scales on the order of the desired resolution. However, since these masks can consist of simple, well-understood patterns this is not foreseen to be a problem.

\section{Analysis}

\begin{figure}[htb]
	\centering
	\includegraphics[width=13.5cm]{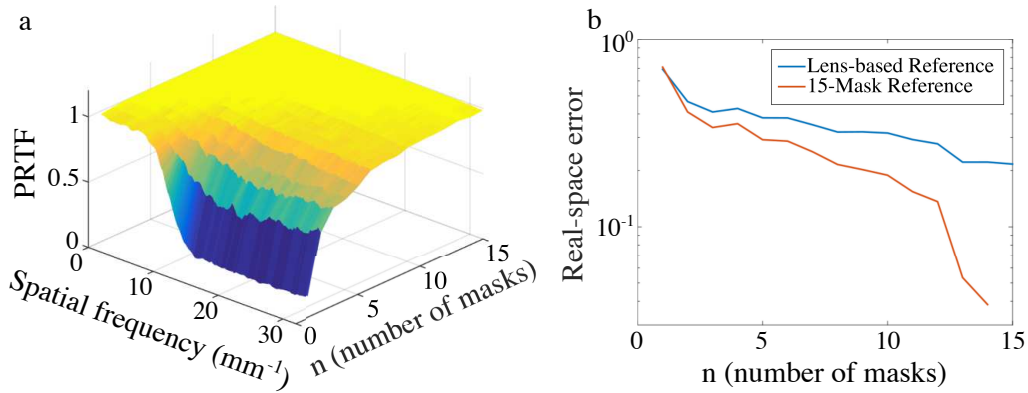}
	\caption{(a) Phase retrieval transfer function as a function of spatial frequency and number of masks used in the reconstruction. (b) Relative error as a function of number of masks, $n$, used in the reconstruction. The error is calculated both relative to the objective, lens-based image and to the reconstructed image for the case when all 15 masks are used in the reconstruction.}
	\label{Analysis}
\end{figure}

A common way to judge reconstruction quality in CDI is through the use of the phase retrieval transfer function, analogous to the modulation transfer function (MTF) that can be measured in more traditional imaging systems \cite{Thibault2006,Chapman2006b}. This curve is a measure of how reliably the phase of the diffraction pattern is reconstructed as a function of spatial frequency \cite{Seaberg2011}. Here we calculate the PRTF as a function of integer number of masks, $n$, used in the reconstruction, ranging from one to fifteen. In order to calculate the PRTF for a given value of $n$, we perform 50 independent reconstructions from random starting guesses, with the average of these reconstructions considered to be the final reconstructed image. Each independent reconstruction is performed using the input-output algorithm. The algorithm is run until the mean square error (MSE) of the Fourier transform of the object relative to the measured diffraction pattern changes by less than one part in $10^6$ from one iteration to the next. The results of the PRTF calculation are shown in Fig.~\ref{Analysis}(a). As can be seen from the figure, good reconstructions are obtained with only 3 or 4 masks, demonstrating stability and uniqueness of the reconstruction. With $n \ge 13$ the phase converges consistently to the same result across all measured spatial frequencies, representing a near-perfect reconstruction of the object.

In order to make an object-space error calculation, the lens-based image is first registered to the 15-mask reconstruction shown in Fig.~\ref{Results}(b) using a least-square fit. The adjustable (global) parameters include coordinate scaling, translation, rotation, and image brightness level. After this image registration, the relative difference between images is calculated between the lens-based image and representative image reconstructions with $1 \le n \le 15$; the resulting curve is shown in Fig.~\ref{Analysis}(b). To ensure consistency, the relative difference between reconstructions with $1 \le n \le 14$ and the 15-mask reconstruction is also shown in Fig.~\ref{Analysis}(b). Both curves imply that the image quality improves with the number of masks used in the reconstruction. Visual inspection of Figs.~\ref{Analysis}(a) and (b) indicates that at high spatial frequencies the PRTF agrees well with the more objective error measurement, implying that in cases where lensless imaging techniques are the only available option this calculation can be trusted to provide a good measure of image quality.

\section{Conclusions}

\label{Conclusions}

The technique demonstrated here takes advantage of randomly coded masks to, in the case of enough masks (typically 3 or 4), guarantee the convergence of phase retrieval algorithms. This is promising for future X-ray application, particularly at X-ray FELs \cite{McNeil2010}. The technique offers similar advantages over basic CDI to those obtained using ptychography. Specifically, there is no need for object-space phase constraints or support constraints. Beyond this, we expect that this technique will be useful for FEL wavefront characterization, as currently available methods either do not provide phase information \cite{Burian2015} or require averaging over many laser shots \cite{Schropp2013}. This technique may also enhance current methods capable of sub-wavelength imaging of terahertz fields with biomedical and metamaterial applications \cite{Blanchard2013a}.

Further development and study of this technique will include investigation into mask position determination using, for instance, a gradient-based search or other methods. This type of position refinement has been shown to be successful in ptychography \cite{Zhang2013a,Tripathi2014}, and would obviate the need for precise alignment of the masks during data acquisition. This type of search is easily parallelizable, meaning reconstruction times should not increase significantly given the proper computer hardware. Additionally, further study on the required number of masks is necessary, given that for dynamic studies at FELs the ideal technique would offer single-shot imaging capability.


%
%

%


\bibliography{RandomMasks}

\begin{thebibliography}{29}%
\makeatletter
\providecommand \@ifxundefined [1]{%
 \@ifx{#1\undefined}
}%
\providecommand \@ifnum [1]{%
 \ifnum #1\expandafter \@firstoftwo
 \else \expandafter \@secondoftwo
 \fi
}%
\providecommand \@ifx [1]{%
 \ifx #1\expandafter \@firstoftwo
 \else \expandafter \@secondoftwo
 \fi
}%
\providecommand \natexlab [1]{#1}%
\providecommand \enquote  [1]{``#1''}%
\providecommand \bibnamefont  [1]{#1}%
\providecommand \bibfnamefont [1]{#1}%
\providecommand \citenamefont [1]{#1}%
\providecommand \href@noop [0]{\@secondoftwo}%
\providecommand \href [0]{\begingroup \@sanitize@url \@href}%
\providecommand \@href[1]{\@@startlink{#1}\@@href}%
\providecommand \@@href[1]{\endgroup#1\@@endlink}%
\providecommand \@sanitize@url [0]{\catcode `\\12\catcode `\$12\catcode
  `\&12\catcode `\#12\catcode `\^12\catcode `\_12\catcode `\%12\relax}%
\providecommand \@@startlink[1]{}%
\providecommand \@@endlink[0]{}%
\providecommand \url  [0]{\begingroup\@sanitize@url \@url }%
\providecommand \@url [1]{\endgroup\@href {#1}{\urlprefix }}%
\providecommand \urlprefix  [0]{URL }%
\providecommand \Eprint [0]{\href }%
\providecommand \doibase [0]{http://dx.doi.org/}%
\providecommand \selectlanguage [0]{\@gobble}%
\providecommand \bibinfo  [0]{\@secondoftwo}%
\providecommand \bibfield  [0]{\@secondoftwo}%
\providecommand \translation [1]{[#1]}%
\providecommand \BibitemOpen [0]{}%
\providecommand \bibitemStop [0]{}%
\providecommand \bibitemNoStop [0]{.\EOS\space}%
\providecommand \EOS [0]{\spacefactor3000\relax}%
\providecommand \BibitemShut  [1]{\csname bibitem#1\endcsname}%
\let\auto@bib@innerbib\@empty
\bibitem [{\citenamefont {Sayre}(1952)}]{Sayre1952}%
  \BibitemOpen
  \bibfield  {author} {\bibinfo {author} {\bibfnamefont {D.}~\bibnamefont
  {Sayre}},\ }\bibfield  {title} {\enquote {\bibinfo {title} {{Some
  implications of a theorem due to Shannon}},}\ }\href {\doibase
  10.1107/S0365110X52002276} {\bibfield  {journal} {\bibinfo  {journal} {Acta
  Crystallographica}\ }\textbf {\bibinfo {volume} {5}},\ \bibinfo {pages} {843}
  (\bibinfo {year} {1952})}\BibitemShut {NoStop}%
\bibitem [{\citenamefont {Miao}, \citenamefont {Charalambous},\ and\
  \citenamefont {Kirz}(1999)}]{Miao1999}%
  \BibitemOpen
  \bibfield  {author} {\bibinfo {author} {\bibfnamefont {J.}~\bibnamefont
  {Miao}}, \bibinfo {author} {\bibfnamefont {P.}~\bibnamefont {Charalambous}},
  \ and\ \bibinfo {author} {\bibfnamefont {J.}~\bibnamefont {Kirz}},\
  }\bibfield  {title} {\enquote {\bibinfo {title} {{Extending the methodology
  of X-ray crystallography to allow imaging of micrometre-sized non-crystalline
  specimens}},}\ }\href {\doibase 10.1038/22498} {\bibfield  {journal}
  {\bibinfo  {journal} {Nature}\ }\textbf {\bibinfo {volume} {400}},\ \bibinfo
  {pages} {342--344} (\bibinfo {year} {1999})}\BibitemShut {NoStop}%
\bibitem [{\citenamefont {Chapman}\ and\ \citenamefont
  {Nugent}(2010)}]{Chapman2010}%
  \BibitemOpen
  \bibfield  {author} {\bibinfo {author} {\bibfnamefont {H.~N.}\ \bibnamefont
  {Chapman}}\ and\ \bibinfo {author} {\bibfnamefont {K.~A.}\ \bibnamefont
  {Nugent}},\ }\bibfield  {title} {\enquote {\bibinfo {title} {{Coherent
  lensless X-ray imaging}},}\ }\href {\doibase 10.1038/nphoton.2010.240}
  {\bibfield  {journal} {\bibinfo  {journal} {Nature Photonics}\ }\textbf
  {\bibinfo {volume} {4}},\ \bibinfo {pages} {833--839} (\bibinfo {year}
  {2010})}\BibitemShut {NoStop}%
\bibitem [{\citenamefont {Miao}\ \emph {et~al.}(2015)\citenamefont {Miao},
  \citenamefont {Ishikawa}, \citenamefont {Robinson},\ and\ \citenamefont
  {Murnane}}]{Miao2015}%
  \BibitemOpen
  \bibfield  {author} {\bibinfo {author} {\bibfnamefont {J.}~\bibnamefont
  {Miao}}, \bibinfo {author} {\bibfnamefont {T.}~\bibnamefont {Ishikawa}},
  \bibinfo {author} {\bibfnamefont {I.~K.}\ \bibnamefont {Robinson}}, \ and\
  \bibinfo {author} {\bibfnamefont {M.~M.}\ \bibnamefont {Murnane}},\
  }\bibfield  {title} {\enquote {\bibinfo {title} {{Beyond crystallography:
  Diffractive imaging using coherent x-ray light sources}},}\ }\href@noop {}
  {\bibfield  {journal} {\bibinfo  {journal} {Science}\ }\textbf {\bibinfo
  {volume} {348}},\ \bibinfo {pages} {530--535} (\bibinfo {year}
  {2015})}\BibitemShut {NoStop}%
\bibitem [{\citenamefont {Thibault}\ \emph {et~al.}(2008)\citenamefont
  {Thibault}, \citenamefont {Dierolf}, \citenamefont {Menzel}, \citenamefont
  {Bunk}, \citenamefont {David},\ and\ \citenamefont
  {Pfeiffer}}]{Thibault2008}%
  \BibitemOpen
  \bibfield  {author} {\bibinfo {author} {\bibfnamefont {P.}~\bibnamefont
  {Thibault}}, \bibinfo {author} {\bibfnamefont {M.}~\bibnamefont {Dierolf}},
  \bibinfo {author} {\bibfnamefont {A.}~\bibnamefont {Menzel}}, \bibinfo
  {author} {\bibfnamefont {O.}~\bibnamefont {Bunk}}, \bibinfo {author}
  {\bibfnamefont {C.}~\bibnamefont {David}}, \ and\ \bibinfo {author}
  {\bibfnamefont {F.}~\bibnamefont {Pfeiffer}},\ }\bibfield  {title} {\enquote
  {\bibinfo {title} {{High-resolution scanning x-ray diffraction
  microscopy}},}\ }\href {\doibase 10.1126/science.1158573} {\bibfield
  {journal} {\bibinfo  {journal} {Science}\ }\textbf {\bibinfo {volume}
  {321}},\ \bibinfo {pages} {379--82} (\bibinfo {year} {2008})}\BibitemShut
  {NoStop}%
\bibitem [{\citenamefont {Maiden}\ and\ \citenamefont
  {Rodenburg}(2009)}]{Maiden2009}%
  \BibitemOpen
  \bibfield  {author} {\bibinfo {author} {\bibfnamefont {A.~M.}\ \bibnamefont
  {Maiden}}\ and\ \bibinfo {author} {\bibfnamefont {J.~M.}\ \bibnamefont
  {Rodenburg}},\ }\bibfield  {title} {\enquote {\bibinfo {title} {{An improved
  ptychographical phase retrieval algorithm for diffractive imaging}},}\ }\href
  {\doibase 10.1016/j.ultramic.2009.05.012} {\bibfield  {journal} {\bibinfo
  {journal} {Ultramicroscopy}\ }\textbf {\bibinfo {volume} {109}},\ \bibinfo
  {pages} {1256--62} (\bibinfo {year} {2009})}\BibitemShut {NoStop}%
\bibitem [{\citenamefont {Wilke}\ \emph {et~al.}(2012)\citenamefont {Wilke},
  \citenamefont {Priebe}, \citenamefont {Bartels}, \citenamefont
  {Giewekemeyer}, \citenamefont {Diaz}, \citenamefont {Karvinen},\ and\
  \citenamefont {Salditt}}]{Wilke2012}%
  \BibitemOpen
  \bibfield  {author} {\bibinfo {author} {\bibfnamefont {R.~N.}\ \bibnamefont
  {Wilke}}, \bibinfo {author} {\bibfnamefont {M.}~\bibnamefont {Priebe}},
  \bibinfo {author} {\bibfnamefont {M.}~\bibnamefont {Bartels}}, \bibinfo
  {author} {\bibfnamefont {K.}~\bibnamefont {Giewekemeyer}}, \bibinfo {author}
  {\bibfnamefont {A.}~\bibnamefont {Diaz}}, \bibinfo {author} {\bibfnamefont
  {P.}~\bibnamefont {Karvinen}}, \ and\ \bibinfo {author} {\bibfnamefont
  {T.}~\bibnamefont {Salditt}},\ }\bibfield  {title} {\enquote {\bibinfo
  {title} {{Hard X-ray imaging of bacterial cells: nano-diffraction and
  ptychographic reconstruction}},}\ }\href {\doibase 10.1364/OE.20.019232}
  {\bibfield  {journal} {\bibinfo  {journal} {Optics Express}\ }\textbf
  {\bibinfo {volume} {20}},\ \bibinfo {pages} {19232} (\bibinfo {year}
  {2012})}\BibitemShut {NoStop}%
\bibitem [{\citenamefont {Hruszkewycz}\ \emph {et~al.}(2013)\citenamefont
  {Hruszkewycz}, \citenamefont {Highland}, \citenamefont {Holt}, \citenamefont
  {Kim}, \citenamefont {Folkman}, \citenamefont {Thompson}, \citenamefont
  {Tripathi}, \citenamefont {Stephenson}, \citenamefont {Hong},\ and\
  \citenamefont {Fuoss}}]{Hruszkewycz2013}%
  \BibitemOpen
  \bibfield  {author} {\bibinfo {author} {\bibfnamefont {S.~O.}\ \bibnamefont
  {Hruszkewycz}}, \bibinfo {author} {\bibfnamefont {M.~J.}\ \bibnamefont
  {Highland}}, \bibinfo {author} {\bibfnamefont {M.~V.}\ \bibnamefont {Holt}},
  \bibinfo {author} {\bibfnamefont {D.}~\bibnamefont {Kim}}, \bibinfo {author}
  {\bibfnamefont {C.~M.}\ \bibnamefont {Folkman}}, \bibinfo {author}
  {\bibfnamefont {C.}~\bibnamefont {Thompson}}, \bibinfo {author}
  {\bibfnamefont {A.}~\bibnamefont {Tripathi}}, \bibinfo {author}
  {\bibfnamefont {G.~B.}\ \bibnamefont {Stephenson}}, \bibinfo {author}
  {\bibfnamefont {S.}~\bibnamefont {Hong}}, \ and\ \bibinfo {author}
  {\bibfnamefont {P.~H.}\ \bibnamefont {Fuoss}},\ }\bibfield  {title} {\enquote
  {\bibinfo {title} {{Imaging local polarization in ferroelectric thin films by
  coherent X-ray bragg projection ptychography}},}\ }\href {\doibase
  10.1103/PhysRevLett.110.177601} {\bibfield  {journal} {\bibinfo  {journal}
  {Physical Review Letters}\ }\textbf {\bibinfo {volume} {110}},\ \bibinfo
  {pages} {1--5} (\bibinfo {year} {2013})}\BibitemShut {NoStop}%
\bibitem [{\citenamefont {Harada}\ \emph {et~al.}(2013)\citenamefont {Harada},
  \citenamefont {Nakasuji}, \citenamefont {Nagata}, \citenamefont {Watanabe},\
  and\ \citenamefont {Kinoshita}}]{Harada2013}%
  \BibitemOpen
  \bibfield  {author} {\bibinfo {author} {\bibfnamefont {T.}~\bibnamefont
  {Harada}}, \bibinfo {author} {\bibfnamefont {M.}~\bibnamefont {Nakasuji}},
  \bibinfo {author} {\bibfnamefont {Y.}~\bibnamefont {Nagata}}, \bibinfo
  {author} {\bibfnamefont {T.}~\bibnamefont {Watanabe}}, \ and\ \bibinfo
  {author} {\bibfnamefont {H.}~\bibnamefont {Kinoshita}},\ }\bibfield  {title}
  {\enquote {\bibinfo {title} {{Phase Imaging of Extreme-Ultraviolet Mask Using
  Coherent Extreme-Ultraviolet Scatterometry Microscope}},}\ }\href {\doibase
  10.7567/JJAP.52.06GB02} {\bibfield  {journal} {\bibinfo  {journal} {Japanese
  Journal of Applied Physics}\ }\textbf {\bibinfo {volume} {52}},\ \bibinfo
  {pages} {06GB02} (\bibinfo {year} {2013})}\BibitemShut {NoStop}%
\bibitem [{\citenamefont {Seaberg}\ \emph {et~al.}(2014)\citenamefont
  {Seaberg}, \citenamefont {Zhang}, \citenamefont {Gardner}, \citenamefont
  {Shanblatt}, \citenamefont {Murnane}, \citenamefont {Kapteyn},\ and\
  \citenamefont {Adams}}]{Seaberg2014}%
  \BibitemOpen
  \bibfield  {author} {\bibinfo {author} {\bibfnamefont {M.~D.}\ \bibnamefont
  {Seaberg}}, \bibinfo {author} {\bibfnamefont {B.}~\bibnamefont {Zhang}},
  \bibinfo {author} {\bibfnamefont {D.~F.}\ \bibnamefont {Gardner}}, \bibinfo
  {author} {\bibfnamefont {E.~R.}\ \bibnamefont {Shanblatt}}, \bibinfo {author}
  {\bibfnamefont {M.~M.}\ \bibnamefont {Murnane}}, \bibinfo {author}
  {\bibfnamefont {H.~C.}\ \bibnamefont {Kapteyn}}, \ and\ \bibinfo {author}
  {\bibfnamefont {D.~E.}\ \bibnamefont {Adams}},\ }\bibfield  {title} {\enquote
  {\bibinfo {title} {{Tabletop nanometer extreme ultraviolet imaging in an
  extended reflection mode using coherent Fresnel ptychography}},}\ }\href@noop
  {} {\bibfield  {journal} {\bibinfo  {journal} {Optica}\ }\textbf {\bibinfo
  {volume} {1}},\ \bibinfo {pages} {39--44} (\bibinfo {year}
  {2014})}\BibitemShut {NoStop}%
\bibitem [{\citenamefont {Maiden}, \citenamefont {Humphry},\ and\ \citenamefont
  {Rodenburg}(2012)}]{Maiden2012a}%
  \BibitemOpen
  \bibfield  {author} {\bibinfo {author} {\bibfnamefont {A.~M.}\ \bibnamefont
  {Maiden}}, \bibinfo {author} {\bibfnamefont {M.~J.}\ \bibnamefont {Humphry}},
  \ and\ \bibinfo {author} {\bibfnamefont {J.~M.}\ \bibnamefont {Rodenburg}},\
  }\bibfield  {title} {\enquote {\bibinfo {title} {{Ptychographic transmission
  microscopy in three dimensions using a multi-slice approach}},}\ }\href
  {http://www.ncbi.nlm.nih.gov/pubmed/23201876} {\bibfield  {journal} {\bibinfo
   {journal} {Journal of the Optical Society of America A, Optics, image
  science, and vision}\ }\textbf {\bibinfo {volume} {29}},\ \bibinfo {pages}
  {1606--14} (\bibinfo {year} {2012})}\BibitemShut {NoStop}%
\bibitem [{\citenamefont {Guizar-Sicairos}\ \emph {et~al.}(2015)\citenamefont
  {Guizar-Sicairos}, \citenamefont {Boon}, \citenamefont {Mader}, \citenamefont
  {Diaz}, \citenamefont {Menzel},\ and\ \citenamefont
  {Bunk}}]{Guizar-Sicairos2015}%
  \BibitemOpen
  \bibfield  {author} {\bibinfo {author} {\bibfnamefont {M.}~\bibnamefont
  {Guizar-Sicairos}}, \bibinfo {author} {\bibfnamefont {J.~J.}\ \bibnamefont
  {Boon}}, \bibinfo {author} {\bibfnamefont {K.}~\bibnamefont {Mader}},
  \bibinfo {author} {\bibfnamefont {A.}~\bibnamefont {Diaz}}, \bibinfo {author}
  {\bibfnamefont {A.}~\bibnamefont {Menzel}}, \ and\ \bibinfo {author}
  {\bibfnamefont {O.}~\bibnamefont {Bunk}},\ }\bibfield  {title} {\enquote
  {\bibinfo {title} {{Quantitative interior x-ray nanotomography by a hybrid
  imaging technique}},}\ }\href@noop {} {\bibfield  {journal} {\bibinfo
  {journal} {Optica}\ }\textbf {\bibinfo {volume} {2}},\ \bibinfo {pages}
  {259--266} (\bibinfo {year} {2015})}\BibitemShut {NoStop}%
\bibitem [{\citenamefont {Cand\`{e}s}, \citenamefont {Li},\ and\ \citenamefont
  {Soltanollotabi}(2013)}]{Candes2013}%
  \BibitemOpen
  \bibfield  {author} {\bibinfo {author} {\bibfnamefont {E.~J.}\ \bibnamefont
  {Cand\`{e}s}}, \bibinfo {author} {\bibfnamefont {X.}~\bibnamefont {Li}}, \
  and\ \bibinfo {author} {\bibfnamefont {M.}~\bibnamefont {Soltanollotabi}},\
  }\bibfield  {title} {\enquote {\bibinfo {title} {{Phase Retrieval from Coded
  Difraction Patterns}},}\ }\href@noop {} {\  (\bibinfo {year} {2013})},\
  \Eprint {http://arxiv.org/abs/1310.3240} {arXiv:1310.3240} \BibitemShut
  {NoStop}%
\bibitem [{\citenamefont {Fogel}, \citenamefont {Waldspurger},\ and\
  \citenamefont {D'Aspremont}(2013)}]{Fogel2013}%
  \BibitemOpen
  \bibfield  {author} {\bibinfo {author} {\bibfnamefont {F.}~\bibnamefont
  {Fogel}}, \bibinfo {author} {\bibfnamefont {I.}~\bibnamefont {Waldspurger}},
  \ and\ \bibinfo {author} {\bibfnamefont {A.}~\bibnamefont {D'Aspremont}},\
  }\bibfield  {title} {\enquote {\bibinfo {title} {{Phase retrieval for imaging
  problems}},}\ }\href {http://arxiv.org/abs/1304.7735} {\  (\bibinfo {year}
  {2013})},\ \Eprint {http://arxiv.org/abs/1304.7735} {arXiv:1304.7735}
  \BibitemShut {NoStop}%
\bibitem [{\citenamefont {Chen}\ and\ \citenamefont
  {Cand\`{e}s}(2015)}]{Chen2015}%
  \BibitemOpen
  \bibfield  {author} {\bibinfo {author} {\bibfnamefont {Y.}~\bibnamefont
  {Chen}}\ and\ \bibinfo {author} {\bibfnamefont {E.~J.}\ \bibnamefont
  {Cand\`{e}s}},\ }\bibfield  {title} {\enquote {\bibinfo {title} {{Solving
  Random Quadratic Systems of Equations Is Nearly as Easy as Solving Linear
  Systems}},}\ }\href@noop {} {\  (\bibinfo {year} {2015})},\ \Eprint
  {http://arxiv.org/abs/1505.05114v1} {arXiv:1505.05114v1} \BibitemShut
  {NoStop}%
\bibitem [{\citenamefont {Waldspurger}, \citenamefont {D'Aspremont},\ and\
  \citenamefont {Mallat}(2015)}]{Waldspurger2015}%
  \BibitemOpen
  \bibfield  {author} {\bibinfo {author} {\bibfnamefont {I.}~\bibnamefont
  {Waldspurger}}, \bibinfo {author} {\bibfnamefont {A.}~\bibnamefont
  {D'Aspremont}}, \ and\ \bibinfo {author} {\bibfnamefont {S.}~\bibnamefont
  {Mallat}},\ }\bibfield  {title} {\enquote {\bibinfo {title} {{Phase recovery,
  MaxCut and complex semidefinite programming}},}\ }\href {\doibase
  10.1007/s10107-013-0738-9} {\bibfield  {journal} {\bibinfo  {journal}
  {Mathematical Programming A}\ }\textbf {\bibinfo {volume} {149}},\ \bibinfo
  {pages} {47--81} (\bibinfo {year} {2015})}\BibitemShut {NoStop}%
\bibitem [{\citenamefont {Shannon}(1949)}]{Shannon1949}%
  \BibitemOpen
  \bibfield  {author} {\bibinfo {author} {\bibfnamefont {C.~E.}\ \bibnamefont
  {Shannon}},\ }\bibfield  {title} {\enquote {\bibinfo {title} {{Communication
  in the Presence of Noise}},}\ }\href {\doibase 10.1109/JPROC.1998.659497}
  {\bibfield  {journal} {\bibinfo  {journal} {Proceedings of the I.R.E.}\
  }\textbf {\bibinfo {volume} {37}},\ \bibinfo {pages} {10--21} (\bibinfo
  {year} {1949})}\BibitemShut {NoStop}%
\bibitem [{\citenamefont {Fienup}(1982)}]{Fienup1982}%
  \BibitemOpen
  \bibfield  {author} {\bibinfo {author} {\bibfnamefont {J.~R.}\ \bibnamefont
  {Fienup}},\ }\bibfield  {title} {\enquote {\bibinfo {title} {{Phase retrieval
  algorithms: a comparison}},}\ }\href
  {http://www.ncbi.nlm.nih.gov/pubmed/20396114} {\bibfield  {journal} {\bibinfo
   {journal} {Applied optics}\ }\textbf {\bibinfo {volume} {21}},\ \bibinfo
  {pages} {2758--69} (\bibinfo {year} {1982})}\BibitemShut {NoStop}%
\bibitem [{\citenamefont {Quiney}\ \emph {et~al.}(2006)\citenamefont {Quiney},
  \citenamefont {Peele}, \citenamefont {Cai}, \citenamefont {Paterson},\ and\
  \citenamefont {Nugent}}]{Quiney2006}%
  \BibitemOpen
  \bibfield  {author} {\bibinfo {author} {\bibfnamefont {H.~M.}\ \bibnamefont
  {Quiney}}, \bibinfo {author} {\bibfnamefont {A.~G.}\ \bibnamefont {Peele}},
  \bibinfo {author} {\bibfnamefont {Z.}~\bibnamefont {Cai}}, \bibinfo {author}
  {\bibfnamefont {D.}~\bibnamefont {Paterson}}, \ and\ \bibinfo {author}
  {\bibfnamefont {K.~A.}\ \bibnamefont {Nugent}},\ }\bibfield  {title}
  {\enquote {\bibinfo {title} {{Diffractive imaging of highly focused X-ray
  fields}},}\ }\href {\doibase 10.1038/nphys218} {\bibfield  {journal}
  {\bibinfo  {journal} {Nature Physics}\ }\textbf {\bibinfo {volume} {2}},\
  \bibinfo {pages} {101--104} (\bibinfo {year} {2006})}\BibitemShut {NoStop}%
\bibitem [{\citenamefont {Abbey}\ \emph {et~al.}(2008)\citenamefont {Abbey},
  \citenamefont {Nugent}, \citenamefont {Williams}, \citenamefont {Clark},
  \citenamefont {Peele}, \citenamefont {Pfeifer}, \citenamefont {de~Jonge},\
  and\ \citenamefont {McNulty}}]{Abbey2008}%
  \BibitemOpen
  \bibfield  {author} {\bibinfo {author} {\bibfnamefont {B.}~\bibnamefont
  {Abbey}}, \bibinfo {author} {\bibfnamefont {K.~A.}\ \bibnamefont {Nugent}},
  \bibinfo {author} {\bibfnamefont {G.~J.}\ \bibnamefont {Williams}}, \bibinfo
  {author} {\bibfnamefont {J.~N.}\ \bibnamefont {Clark}}, \bibinfo {author}
  {\bibfnamefont {A.~G.}\ \bibnamefont {Peele}}, \bibinfo {author}
  {\bibfnamefont {M.~A.}\ \bibnamefont {Pfeifer}}, \bibinfo {author}
  {\bibfnamefont {M.}~\bibnamefont {de~Jonge}}, \ and\ \bibinfo {author}
  {\bibfnamefont {I.}~\bibnamefont {McNulty}},\ }\bibfield  {title} {\enquote
  {\bibinfo {title} {{Keyhole coherent diffractive imaging}},}\ }\href
  {\doibase 10.1038/nphys896} {\bibfield  {journal} {\bibinfo  {journal}
  {Nature Physics}\ }\textbf {\bibinfo {volume} {4}},\ \bibinfo {pages}
  {394--398} (\bibinfo {year} {2008})}\BibitemShut {NoStop}%
\bibitem [{\citenamefont {Thibault}\ \emph {et~al.}(2006)\citenamefont
  {Thibault}, \citenamefont {Elser}, \citenamefont {Jacobsen}, \citenamefont
  {Shapiro},\ and\ \citenamefont {Sayre}}]{Thibault2006}%
  \BibitemOpen
  \bibfield  {author} {\bibinfo {author} {\bibfnamefont {P.}~\bibnamefont
  {Thibault}}, \bibinfo {author} {\bibfnamefont {V.}~\bibnamefont {Elser}},
  \bibinfo {author} {\bibfnamefont {C.}~\bibnamefont {Jacobsen}}, \bibinfo
  {author} {\bibfnamefont {D.}~\bibnamefont {Shapiro}}, \ and\ \bibinfo
  {author} {\bibfnamefont {D.}~\bibnamefont {Sayre}},\ }\bibfield  {title}
  {\enquote {\bibinfo {title} {{Reconstruction of a yeast cell from X-ray
  diffraction data}},}\ }\href {\doibase 10.1107/S0108767306016515} {\bibfield
  {journal} {\bibinfo  {journal} {Acta Crystallographica Section A: Foundations
  of Crystallography}\ }\textbf {\bibinfo {volume} {62}},\ \bibinfo {pages}
  {248--261} (\bibinfo {year} {2006})}\BibitemShut {NoStop}%
\bibitem [{\citenamefont {Chapman}\ \emph {et~al.}(2006)\citenamefont
  {Chapman}, \citenamefont {Barty}, \citenamefont {Marchesini}, \citenamefont
  {Noy}, \citenamefont {Hau-riege}, \citenamefont {Cui}, \citenamefont
  {Howells}, \citenamefont {Rosen}, \citenamefont {He}, \citenamefont {Spence},
  \citenamefont {Weierstall}, \citenamefont {Beetz}, \citenamefont {Jacobsen},\
  and\ \citenamefont {Shapiro}}]{Chapman2006b}%
  \BibitemOpen
  \bibfield  {author} {\bibinfo {author} {\bibfnamefont {H.~N.}\ \bibnamefont
  {Chapman}}, \bibinfo {author} {\bibfnamefont {A.}~\bibnamefont {Barty}},
  \bibinfo {author} {\bibfnamefont {S.}~\bibnamefont {Marchesini}}, \bibinfo
  {author} {\bibfnamefont {A.}~\bibnamefont {Noy}}, \bibinfo {author}
  {\bibfnamefont {S.~P.}\ \bibnamefont {Hau-riege}}, \bibinfo {author}
  {\bibfnamefont {C.}~\bibnamefont {Cui}}, \bibinfo {author} {\bibfnamefont
  {M.~R.}\ \bibnamefont {Howells}}, \bibinfo {author} {\bibfnamefont
  {R.}~\bibnamefont {Rosen}}, \bibinfo {author} {\bibfnamefont
  {H.}~\bibnamefont {He}}, \bibinfo {author} {\bibfnamefont {J.~C.~H.}\
  \bibnamefont {Spence}}, \bibinfo {author} {\bibfnamefont {U.}~\bibnamefont
  {Weierstall}}, \bibinfo {author} {\bibfnamefont {T.}~\bibnamefont {Beetz}},
  \bibinfo {author} {\bibfnamefont {C.}~\bibnamefont {Jacobsen}}, \ and\
  \bibinfo {author} {\bibfnamefont {D.}~\bibnamefont {Shapiro}},\ }\bibfield
  {title} {\enquote {\bibinfo {title} {{High-resolution ab initio
  three-dimensional x-ray diffraction microscopy}},}\ }\href@noop {} {\bibfield
   {journal} {\bibinfo  {journal} {J. Opt. Soc. Am. A}\ }\textbf {\bibinfo
  {volume} {23}},\ \bibinfo {pages} {1179--1200} (\bibinfo {year}
  {2006})}\BibitemShut {NoStop}%
\bibitem [{\citenamefont {Seaberg}\ \emph {et~al.}(2011)\citenamefont
  {Seaberg}, \citenamefont {Adams}, \citenamefont {Townsend}, \citenamefont
  {Raymondson}, \citenamefont {Schlotter}, \citenamefont {Liu}, \citenamefont
  {Menoni}, \citenamefont {Rong}, \citenamefont {Chen}, \citenamefont {Miao},
  \citenamefont {Kapteyn},\ and\ \citenamefont {Murnane}}]{Seaberg2011}%
  \BibitemOpen
  \bibfield  {author} {\bibinfo {author} {\bibfnamefont {M.~D.}\ \bibnamefont
  {Seaberg}}, \bibinfo {author} {\bibfnamefont {D.~E.}\ \bibnamefont {Adams}},
  \bibinfo {author} {\bibfnamefont {E.~L.}\ \bibnamefont {Townsend}}, \bibinfo
  {author} {\bibfnamefont {D.~A.}\ \bibnamefont {Raymondson}}, \bibinfo
  {author} {\bibfnamefont {W.~F.}\ \bibnamefont {Schlotter}}, \bibinfo {author}
  {\bibfnamefont {Y.}~\bibnamefont {Liu}}, \bibinfo {author} {\bibfnamefont
  {C.~S.}\ \bibnamefont {Menoni}}, \bibinfo {author} {\bibfnamefont
  {L.}~\bibnamefont {Rong}}, \bibinfo {author} {\bibfnamefont {C.-C.}\
  \bibnamefont {Chen}}, \bibinfo {author} {\bibfnamefont {J.}~\bibnamefont
  {Miao}}, \bibinfo {author} {\bibfnamefont {H.~C.}\ \bibnamefont {Kapteyn}}, \
  and\ \bibinfo {author} {\bibfnamefont {M.~M.}\ \bibnamefont {Murnane}},\
  }\bibfield  {title} {\enquote {\bibinfo {title} {{Ultrahigh 22 nm resolution
  coherent diffractive imaging using a desktop 13 nm high harmonic source}},}\
  }\href {\doibase 10.1364/OE.19.022470} {\bibfield  {journal} {\bibinfo
  {journal} {Optics express}\ }\textbf {\bibinfo {volume} {19}},\ \bibinfo
  {pages} {22470--9} (\bibinfo {year} {2011})}\BibitemShut {NoStop}%
\bibitem [{\citenamefont {McNeil}\ and\ \citenamefont
  {Thompson}(2010)}]{McNeil2010}%
  \BibitemOpen
  \bibfield  {author} {\bibinfo {author} {\bibfnamefont {B.~W.~J.}\
  \bibnamefont {McNeil}}\ and\ \bibinfo {author} {\bibfnamefont {N.~R.}\
  \bibnamefont {Thompson}},\ }\bibfield  {title} {\enquote {\bibinfo {title}
  {{X-ray free-electron lasers}},}\ }\href {\doibase 10.1038/nphoton.2010.239}
  {\bibfield  {journal} {\bibinfo  {journal} {Nature Photonics}\ }\textbf
  {\bibinfo {volume} {4}},\ \bibinfo {pages} {814--821} (\bibinfo {year}
  {2010})}\BibitemShut {NoStop}%
\bibitem [{\citenamefont {Burian}\ \emph {et~al.}(2015)\citenamefont {Burian},
  \citenamefont {H\'{a}jkov\'{a}}, \citenamefont {Chalupsk\'{y}}, \citenamefont
  {Vy\v{s}\'{\i}n}, \citenamefont {Boh\'{a}\v{c}ek}, \citenamefont
  {Pře\v{c}ek}, \citenamefont {Wild}, \citenamefont {\"{O}zkan}, \citenamefont
  {Coppola}, \citenamefont {Farahani}, \citenamefont {Schulz}, \citenamefont
  {Sinn}, \citenamefont {Tschentscher}, \citenamefont {Gaudin}, \citenamefont
  {Bajt}, \citenamefont {Tiedtke}, \citenamefont {Toleikis}, \citenamefont
  {Chapman}, \citenamefont {Loch}, \citenamefont {Jurek}, \citenamefont
  {Sobierajski}, \citenamefont {Krzywinski}, \citenamefont {Moeller},
  \citenamefont {Harmand}, \citenamefont {Galasso}, \citenamefont {Nagasono},
  \citenamefont {Saskl}, \citenamefont {Sov\'{a}k},\ and\ \citenamefont
  {Juha}}]{Burian2015}%
  \BibitemOpen
  \bibfield  {author} {\bibinfo {author} {\bibfnamefont {T.}~\bibnamefont
  {Burian}}, \bibinfo {author} {\bibfnamefont {V.}~\bibnamefont
  {H\'{a}jkov\'{a}}}, \bibinfo {author} {\bibfnamefont {J.}~\bibnamefont
  {Chalupsk\'{y}}}, \bibinfo {author} {\bibfnamefont {L.}~\bibnamefont
  {Vy\v{s}\'{\i}n}}, \bibinfo {author} {\bibfnamefont {P.}~\bibnamefont
  {Boh\'{a}\v{c}ek}}, \bibinfo {author} {\bibfnamefont {M.}~\bibnamefont
  {Pře\v{c}ek}}, \bibinfo {author} {\bibfnamefont {J.}~\bibnamefont {Wild}},
  \bibinfo {author} {\bibfnamefont {C.}~\bibnamefont {\"{O}zkan}}, \bibinfo
  {author} {\bibfnamefont {N.}~\bibnamefont {Coppola}}, \bibinfo {author}
  {\bibfnamefont {S.~D.}\ \bibnamefont {Farahani}}, \bibinfo {author}
  {\bibfnamefont {J.}~\bibnamefont {Schulz}}, \bibinfo {author} {\bibfnamefont
  {H.}~\bibnamefont {Sinn}}, \bibinfo {author} {\bibfnamefont {T.}~\bibnamefont
  {Tschentscher}}, \bibinfo {author} {\bibfnamefont {J.}~\bibnamefont
  {Gaudin}}, \bibinfo {author} {\bibfnamefont {S.}~\bibnamefont {Bajt}},
  \bibinfo {author} {\bibfnamefont {K.}~\bibnamefont {Tiedtke}}, \bibinfo
  {author} {\bibfnamefont {S.}~\bibnamefont {Toleikis}}, \bibinfo {author}
  {\bibfnamefont {H.~N.}\ \bibnamefont {Chapman}}, \bibinfo {author}
  {\bibfnamefont {R.~A.}\ \bibnamefont {Loch}}, \bibinfo {author}
  {\bibfnamefont {M.}~\bibnamefont {Jurek}}, \bibinfo {author} {\bibfnamefont
  {R.}~\bibnamefont {Sobierajski}}, \bibinfo {author} {\bibfnamefont
  {J.}~\bibnamefont {Krzywinski}}, \bibinfo {author} {\bibfnamefont
  {S.}~\bibnamefont {Moeller}}, \bibinfo {author} {\bibfnamefont
  {M.}~\bibnamefont {Harmand}}, \bibinfo {author} {\bibfnamefont
  {G.}~\bibnamefont {Galasso}}, \bibinfo {author} {\bibfnamefont
  {M.}~\bibnamefont {Nagasono}}, \bibinfo {author} {\bibfnamefont
  {K.}~\bibnamefont {Saskl}}, \bibinfo {author} {\bibfnamefont
  {P.}~\bibnamefont {Sov\'{a}k}}, \ and\ \bibinfo {author} {\bibfnamefont
  {L.}~\bibnamefont {Juha}},\ }\bibfield  {title} {\enquote {\bibinfo {title}
  {{Soft x-ray free-electron laser induced damage to inorganic
  scintillators}},}\ }\href {\doibase 10.1364/OME.5.000254} {\bibfield
  {journal} {\bibinfo  {journal} {Optical Materials Express}\ }\textbf
  {\bibinfo {volume} {5}},\ \bibinfo {pages} {254} (\bibinfo {year}
  {2015})}\BibitemShut {NoStop}%
\bibitem [{\citenamefont {Schropp}\ \emph {et~al.}(2013)\citenamefont
  {Schropp}, \citenamefont {Hoppe}, \citenamefont {Meier}, \citenamefont
  {Patommel}, \citenamefont {Seiboth}, \citenamefont {Lee}, \citenamefont
  {Nagler}, \citenamefont {Galtier}, \citenamefont {Arnold}, \citenamefont
  {Zastrau}, \citenamefont {Hastings}, \citenamefont {Nilsson}, \citenamefont
  {Uhl\'{e}n}, \citenamefont {Vogt}, \citenamefont {Hertz},\ and\ \citenamefont
  {Schroer}}]{Schropp2013}%
  \BibitemOpen
  \bibfield  {author} {\bibinfo {author} {\bibfnamefont {A.}~\bibnamefont
  {Schropp}}, \bibinfo {author} {\bibfnamefont {R.}~\bibnamefont {Hoppe}},
  \bibinfo {author} {\bibfnamefont {V.}~\bibnamefont {Meier}}, \bibinfo
  {author} {\bibfnamefont {J.}~\bibnamefont {Patommel}}, \bibinfo {author}
  {\bibfnamefont {F.}~\bibnamefont {Seiboth}}, \bibinfo {author} {\bibfnamefont
  {H.~J.}\ \bibnamefont {Lee}}, \bibinfo {author} {\bibfnamefont
  {B.}~\bibnamefont {Nagler}}, \bibinfo {author} {\bibfnamefont {E.~C.}\
  \bibnamefont {Galtier}}, \bibinfo {author} {\bibfnamefont {B.}~\bibnamefont
  {Arnold}}, \bibinfo {author} {\bibfnamefont {U.}~\bibnamefont {Zastrau}},
  \bibinfo {author} {\bibfnamefont {J.~B.}\ \bibnamefont {Hastings}}, \bibinfo
  {author} {\bibfnamefont {D.}~\bibnamefont {Nilsson}}, \bibinfo {author}
  {\bibfnamefont {F.}~\bibnamefont {Uhl\'{e}n}}, \bibinfo {author}
  {\bibfnamefont {U.}~\bibnamefont {Vogt}}, \bibinfo {author} {\bibfnamefont
  {H.~M.}\ \bibnamefont {Hertz}}, \ and\ \bibinfo {author} {\bibfnamefont
  {C.~G.}\ \bibnamefont {Schroer}},\ }\bibfield  {title} {\enquote {\bibinfo
  {title} {{Full spatial characterization of a nanofocused x-ray free-electron
  laser beam by ptychographic imaging}},}\ }\href {\doibase 10.1038/srep01633}
  {\bibfield  {journal} {\bibinfo  {journal} {Scientific reports}\ }\textbf
  {\bibinfo {volume} {3}},\ \bibinfo {pages} {1633} (\bibinfo {year}
  {2013})}\BibitemShut {NoStop}%
\bibitem [{\citenamefont {Blanchard}\ \emph {et~al.}(2013)\citenamefont
  {Blanchard}, \citenamefont {Doi}, \citenamefont {Tanaka},\ and\ \citenamefont
  {Tanaka}}]{Blanchard2013a}%
  \BibitemOpen
  \bibfield  {author} {\bibinfo {author} {\bibfnamefont {F.}~\bibnamefont
  {Blanchard}}, \bibinfo {author} {\bibfnamefont {A.}~\bibnamefont {Doi}},
  \bibinfo {author} {\bibfnamefont {T.}~\bibnamefont {Tanaka}}, \ and\ \bibinfo
  {author} {\bibfnamefont {K.}~\bibnamefont {Tanaka}},\ }\bibfield  {title}
  {\enquote {\bibinfo {title} {{Real-Time, Subwavelength Terahertz Imaging}},}\
  }\href {\doibase 10.1146/annurev-matsci-071312-121656} {\bibfield  {journal}
  {\bibinfo  {journal} {Annual Review of Materials Research}\ }\textbf
  {\bibinfo {volume} {43}},\ \bibinfo {pages} {237--259} (\bibinfo {year}
  {2013})}\BibitemShut {NoStop}%
\bibitem [{\citenamefont {Zhang}\ \emph {et~al.}(2013)\citenamefont {Zhang},
  \citenamefont {Peterson}, \citenamefont {Vila-Comamala}, \citenamefont
  {Diaz}, \citenamefont {Bean}, \citenamefont {Chen}, \citenamefont {Menzel},
  \citenamefont {Robinson},\ and\ \citenamefont {Rodenburg}}]{Zhang2013a}%
  \BibitemOpen
  \bibfield  {author} {\bibinfo {author} {\bibfnamefont {F.}~\bibnamefont
  {Zhang}}, \bibinfo {author} {\bibfnamefont {I.}~\bibnamefont {Peterson}},
  \bibinfo {author} {\bibfnamefont {J.}~\bibnamefont {Vila-Comamala}}, \bibinfo
  {author} {\bibfnamefont {A.}~\bibnamefont {Diaz}}, \bibinfo {author}
  {\bibfnamefont {R.}~\bibnamefont {Bean}}, \bibinfo {author} {\bibfnamefont
  {B.}~\bibnamefont {Chen}}, \bibinfo {author} {\bibfnamefont {A.}~\bibnamefont
  {Menzel}}, \bibinfo {author} {\bibfnamefont {I.~K.}\ \bibnamefont
  {Robinson}}, \ and\ \bibinfo {author} {\bibfnamefont {J.~M.}\ \bibnamefont
  {Rodenburg}},\ }\bibfield  {title} {\enquote {\bibinfo {title} {{Translation
  position determination in ptychographic coherent diffraction imaging}},}\
  }\href {\doibase 10.1364/OE.21.013592} {\bibfield  {journal} {\bibinfo
  {journal} {Optics express}\ }\textbf {\bibinfo {volume} {21}},\ \bibinfo
  {pages} {13592--13606} (\bibinfo {year} {2013})}\BibitemShut {NoStop}%
\bibitem [{\citenamefont {Tripathi}, \citenamefont {Mcnulty},\ and\
  \citenamefont {Shpyrko}(2014)}]{Tripathi2014}%
  \BibitemOpen
  \bibfield  {author} {\bibinfo {author} {\bibfnamefont {A.}~\bibnamefont
  {Tripathi}}, \bibinfo {author} {\bibfnamefont {I.}~\bibnamefont {Mcnulty}}, \
  and\ \bibinfo {author} {\bibfnamefont {O.~G.}\ \bibnamefont {Shpyrko}},\
  }\bibfield  {title} {\enquote {\bibinfo {title} {{Ptychographic overlap
  constraint errors and the limits of their numerical recovery using conjugate
  gradient descent methods}},}\ }\href {\doibase 10.1364/OE.22.001452}
  {\bibfield  {journal} {\bibinfo  {journal} {Optics Express}\ }\textbf
  {\bibinfo {volume} {22}},\ \bibinfo {pages} {1452--1466} (\bibinfo {year}
  {2014})}\BibitemShut {NoStop}%
\end{thebibliography}%

\end{document}